\documentclass[conference]{IEEEtran}
\usepackage{cite}
\usepackage{amsmath,amssymb,amsfonts}
\usepackage{algorithmic}
\usepackage{graphicx}
\usepackage{textcomp}
\usepackage{xcolor}
\usepackage{enumitem}
\setlist{topsep=0pt, leftmargin=*}

\makeatletter
 \let\old@ps@headings\ps@headings
 \let\old@ps@IEEEtitlepagestyle\ps@IEEEtitlepagestyle
 \def\confheader#1{%
 \def\ps@headings{%
 \old@ps@headings%
 \def\@oddhead{\strut\hfill#1\hfill\strut}%
 \def\@evenhead{\strut\hfill#1\hfill\strut}%
 }%
 \def\ps@IEEEtitlepagestyle{%
 \old@ps@IEEEtitlepagestyle%
 \def\@oddhead{\strut\hfill#1\hfill\strut}%
 \def\@evenhead{\strut\hfill#1\hfill\strut}%
 }%
 \ps@headings%
 }
 \makeatother
 
\usepackage{hyperref}

\begin{document}

\title{Intent-based Network Management and Orchestration for Smart Distribution Grids\\
}
\author{
\IEEEauthorblockN{
Kashif Mehmood,
H. V. Kalpanie Mendis,
Katina Kralevska,
and Poul E. Heegaard
}
\IEEEauthorblockA{Department of Information Security and Communication Technology,\\
NTNU - Norwegian University of Science and Technology, N-7491 Trondheim, Norway \\  
Email: \{kashif.mehmood; kalpanie.mendis; katinak; poul.heegaard\}@ntnu.no}}

\maketitle

\begin{abstract}
5G technology complements the enabling of communication services for different vertical industries such as smart distribution grids. Automation is an integral and necessary part of the power distribution grid operation and management. This paper postulates a framework by which the smart distribution grid can obtain service-oriented communication services using 5G network slicing and intent-based networking (IBN). IBN provides an interface to service users and network stakeholders to cooperate through a high level abstraction model of service provisioning in a network agnostic manner. The automation and adaptability of the distribution grid are facilitated by using the dynamic and closed-loop mechanism of IBN together with network slicing and network function virtualization for network management and orchestration. We identify the automation parts of the power distribution grid and illustrate the intent processing and its inclusion in the definition of network slice instances, service and network configuration models. 
\end{abstract}


\section{Introduction}\label{sec:intro}
Network virtualization has transformed the design, deployment, control, management and orchestration of network components.
Next-generation communication networks such as 5G, with their programmable nature, enable creating and managing customized networks (slices) triggered by on-demand service requests \cite{9165419}.

Intent-based networking (IBN) is a novel technology for automating and orchestrating networks that can quickly adapt to ever-changing applications and services \cite{intent-defs}. Intent is a declaration of operational goals that a network should meet and outcomes that the network is supposed to deliver, without specifying how to achieve them. IBN management provides the ability to couple services and network functions to offer on-demand and swift service to various verticals with requirements (functionality, quality-of-service (QoS)) expressed by the stakeholders in a vertical \cite{5GPP-Verticals-Whitepaper}. User intents help in determining the scope of the requested services, and an intent controller extracts the required service requirements from the intent representation and established service level agreements (SLAs).  An intent-based network is an intelligent network that can automatically convert, deploy, and configure network resources according to the users' intents. A closed-loop feedback mechanism provides the ability to continuously monitor and adapt the established services to network conditions and stakeholders' intents during orchestration life-cycle.

5G and beyond networks have enabled the application of information and communication technologies (ICT) in a diverse set of verticals and industries, and one of them is power distribution. Network providers in smart grids can utilize evolving 5G, network slicing, and IBN frameworks to provide differentiated services, efficient distribution and power consumption, reduced transmission loss, improved reliability of the energy grid operations to the customers, reduction of human intervention, and reduced operational cost \cite{SGrid-sruvey1}.\par 

The service requirements for the smart grids are based upon several possible application scenarios stated as but not restricted to 1) distribution system monitoring and control, 2) distribution system protection, 3) management of distributed energy sources (DERs), and 4) demand response. These application use cases require the utilization of diverse services offered by the 5G networks due to their heterogeneous performance metrics thus, belong to the three main use case categories: enhanced mobile broadband (eMBB), ultra-reliable and low latency communications (URLLC), and massive machine-type communications (mMTC). For example, feeder automation requires latency in the order of milliseconds with very high reliability of 99.999\% (URLLC), and the distributed nature of power generation and other grid units requires massive access support for a large number of devices simultaneously (mMTC) \cite{SGrid-Demanding}. Hence, the provisioning of differentiated services to the smart distribution grid requires accurate and timely orchestration and automation of the necessary network functions and resources. The inclusion of user intents in the management of the smart distribution grid communication provides vast benefits of cost reduction, enhanced power reliability and quality, less power outages, improved grid resiliency, safety and security to the different stakeholders.

The application of IBN for different industrial use cases, such as protection paths in ring networks and data dissemination in the vehicular edge computing ecosystem, has been investigated in \cite{10.1145/3243318.3243324} and \cite{9123601}, respectively. A recent study \cite{8893119} focuses on the utilization of higher-level instructions in the form of intents to perform configuration and management of 5G networks in a cross-platform environment. The work in \cite{8645632} reinforces the premise of IBN management as a promising direction in solving the connectivity and configuration issues in smart grids as well as making the network resilient.\par

In this paper, we apply the concept of intent-based network management in the domain of smart distribution grids. Our contributions are summarized as follows:
\begin{itemize}
     \item An intent and service orchestration mechanism is devised for intent translation and creation of slice instances, service delivery and network configuration models.
     \item Identification of means of automation for the smart grid use case with a discussion of different stakeholders and how IBN can help in automating dynamic communication service life-cycle.
    \item We propose an intent-based service management and orchestration framework that incorporates the smart grid customers' intents and the SLAs with the service providers and network operators, reflected in the obtained service models.
    \item Smart grid vertical example is provided to identify the service design and orchestration process using network slicing and YANG service model.
\end{itemize}

The remainder of this paper is organized as follows. Sec.~\ref{sec:ibn} provides an overview of IBN management and orchestration. Sec.~\ref{sec:grid} provides a description of the smart grid use case. Sec.~\ref{sec:architect} presents the proposed intent-based network management architecture for smart grids. Sec.~\ref{sec:con} concludes the paper.
\begin{figure*}[t!]
    \includegraphics[width=1\textwidth]{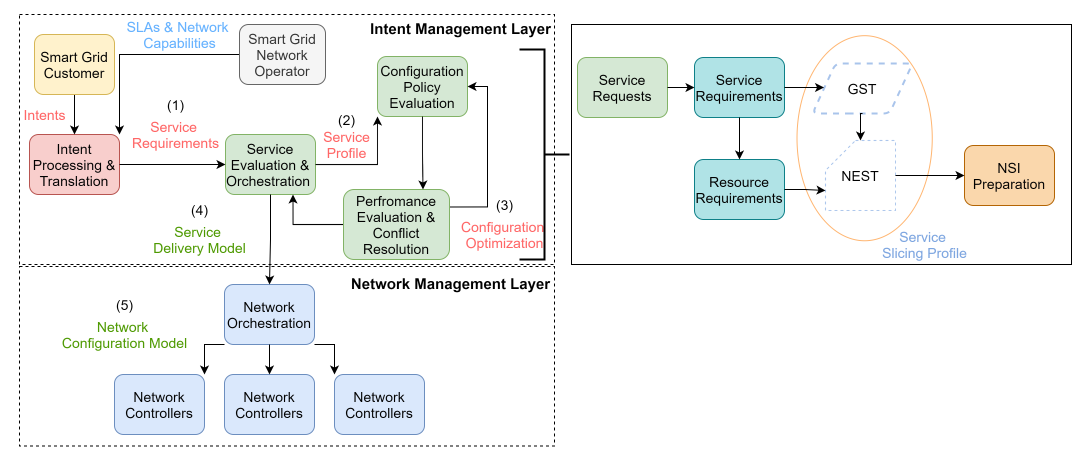}
	\caption{Intent processing and translation into slicing profiles, service delivery models, and network configuration models.}	\label{image:intents}
\end{figure*}
\section{Intent-based Network Management and Orchestration}\label{sec:ibn}
The service-based architecture of 5G networks supports the provisioning of communication services and network infrastructures based on differentiated service models. 
The ultimate goal of IBN management is to provide a generalized input from the stakeholders to the management plane in order to orchestrate efficiently and provision resources for different services.\par
\subsection{Intents: Definition, Life-cycle and Network Integration}
An intent can be defined as an \textit{abstract, high-level policy used to operate a network} with different types of services based on the feedback received from the intent and network management layers \cite{Aut-Net}. Intents are different from other policy-based user inputs in the sense that they provide a general direction about the objectives of the intents of the users like the service customers. The network must understand these objectives and implement the necessary configurations and technologies to satisfy the intents. The orchestration of the required resources can be done by the network orchestrators in order to deliver the desired service and user key performance indicators (KPIs) such as latency, availability and security. Intents are integrated into the service design process and augment an automated intent processing mechanism to convert abstracted user requirements into service and resource provisioning configurations for the underlying network orchestrators. The network creates a service model comprising of a customer service model and a service delivery model from the translated user requirements acquired through the customer intents \cite{service-model}. Network function virtualization (NFV) has enabled the automation in the deployment of required network functions for different service models in a service-centric network environment.

\subsection{From Intents to Service and Network Configuration Models}
Intents are converted into a form which the network and service orchestrators can use to meet end-to-end (E2E) customer performance requirements. In an IBN design, the service model is obtained after the processing of user intents along with the consideration of relevant SLAs from the network operators, as presented in Fig.\ref{image:intents}. 

A service model defines the service provided to the user in a network environment. The intents are translated into a set of service requirements for each user, step (1) in Fig.\ref{image:intents}, and converted into different available service profiles supported by the network, step (2), \cite{intent-defs}. A smart grid user can require different services according to the application use cases explained in Section \ref{sec:grid}. The service profiles are optimized and rigorously evaluated for potential trade-offs and conflicts of interest in a closed-loop system consisting of service, relevant configuration, and performance evaluation as per required resources, step (3). The service delivery models are forwarded to network orchestration, step (4), consisting of optimized configuration guidelines in order to meet the relevant service requirements. The network orchestrator generates the network configurations for different requested services and assists the network controllers in the deployment of relevant network functions, step (5).

\subsection{From Intents and Service Requests to Slice Instances}
Various organizations have provided frameworks for orchestration and implementation of communication service requirements in 5G networks. GSMA utilizes the concept of generic network slice templates (GST) to specify the communication service requirements \cite{GSMA-GST}. Network slicing is utilized to orchestrate resources by the slicing orchestrator through the management and network orchestrator (MANO). Intent orchestrator provides the required set of service requirements for the user intent after validating with the agreed SLA. GST is network deployment agnostic representation of a slicing module with specified slice type acting as one of the several  network slice type  (NEST). Several network slice instances (NSIs) can be generated to fulfill service requirements from a single customer depending upon the propagated service requests \cite{ETSI-5GArch}. 
A network slice provider accepts the NEST and triggers the network slice instantiation process in order to perform necessary resource allocations from the network management layer as depicted in Fig. \ref{image:intents}.\par
The slice orchestrator is responsible for the instantiation of different NSIs in order to fulfill the requested service. This is accomplished through the definition of network function (NF) service chains by the MANO framework. The process is completed in accordance with the service requirements and mapped onto user intents through the intent and service orchestrators. A key consideration here is the ability to modify the slice life-cycle through the slicing orchestrator in order to address the dynamic nature of the grid environment and changing user intents accordingly. 

\subsection{Automation in Intent Processing and Service Design}
Automation in service and network orchestration requires a dynamic framework to include service users in the service design process. A closed optimization loop exists between an intent monitor module and the intent orchestrator in order to achieve a dynamic and user intent-based service life-cycle management. The intent processing utilizes language processing mechanisms to convert the high-level network agnostic intents into different service profiles and service-specific configurations to be provided to the network elements in the network management layer. The definition of service objectives and profiles requires exposure of network capabilities and SLAs agreed upon between the customer and the service providers and between the other stakeholders \cite{Aut-Net}.
The automation required during the intent processing and service design process itself can be achieved through a policy based optimization framework involving machine learning algorithms. The intent and service orchestrators utilize local databases to store different predefined service and intent templates in order to map and adapt different network agnostic intent translation and service mapping policies. The required data for the learning based automation consists of the consumer groups, service provider and operator SLAs, service slicing templates for eMBB, mMTC and uRLLC and the monitoring data from the smart grid network.

\section{Power Distribution Grid Automation}\label{sec:grid}
The well-functioning of modern society is highly dependent on the reliability and resiliency of the power distribution grid, which plays a key role as a critical infrastructure. As more distributed energy resources (DERs) are integrated at the low and medium voltage levels, the flexibility of the power supply decreases while hindering the reliability of the power grid. The next-generation power distribution grid or the smart distribution grid is forecasted to be more flexible and resilient with increased power consumption, local production, and storage options with added communication capabilities, control, and intelligence.
%



\begin{table*}[tb]
  \centering \small
  \caption{Service requirements of distribution grid operations for communication networks \cite{mendis20195g}.}
  \label{table:1}

  \begin{tabular}{ llllll }

    DG operation & 5G use case & Reliability & Latency & Bandwidth  & \#Devices \\

    \hline
    WAMS & URLLC  & high & low & low & low \\
    Protection, FLISR & URLLC  & high & low & low & low \\
    AMI & mMTC  & low & high & low & high \\
    Remote inspection  & eMBB  & low & low & high  & low\\
    \hline
  \end{tabular}
\end{table*}

\subsection{Main power distribution grid operations}  
Smart grid provides monitoring, protection, and optimization functions automatically to the operation of the interconnected elements covering generation, transmission, distribution, and to the consumers \cite{yan2012survey}. 
It can intelligently integrate the needs and capabilities of all grid stakeholders connected to it \cite{6112018}.
 Power distribution automation plays a crucial role in smart grids. Distribution automation is the path towards a self healing grid. A self healing grid is \textit{a system that uses information, sensing, control and communication technologies to allow it to deal with unforeseen events and minimize their adverse impact}. As the grid dynamics increase and more closed-loop controls are implemented, communication networks become more integral to the smart distribution grid operations. With the enhanced real-time monitoring and controlling capability provided by the deployment of new sensors, digital measuring equipment, and automated protection relays, there is a need for a low latency, high reliable, real-time, and interactive communications system.

\subsubsection{Distribution system monitoring and control} Smart monitoring and control are essential elements of the smart distribution grid. Smart operation of distribution grid requires several measurements including voltage, current, frequency, power flows and equipment status information. 

Several ICT-based applications are used to monitor the status of the power distribution grid. The distribution management system (DMS) is capable of collecting, and analyzing real-time information on the distribution grid. Existing operation and (remote) control of the power grid heavily rely on the supervisory control and data acquisition (SCADA) system. It mainly handles the local and remote operation of the high voltage equipment, information and measurement acquisition, and emergency events (alarms). Phasor measurement units (PMUs) obtain synchronized measurements across a wide area of the power grid and constitute the wide area measuring system (WAMS). WAMS aims to detect power grid disturbances in real-time using state estimation algorithms with extreme reliability and ultra-low latency. Intelligent electronic devices (IEDs) are used for protection purposes, and use IEC 61850 protocol to communicate with other IEDs. We assume that the technology and cost development allow the deployment of PMUs in the future distribution grid, which are mostly deployed in the transmission grid today. 

\subsubsection{Distribution system protection} A failure in the power distribution grid is stressful. Power outages have to be handled in milliseconds for critical power consumers. Thus, self-healing solutions empowered by automated switching, fault location, isolation and service restoration (FLISR) are vital. Several automated protection functions in the distribution grid require a time-critical exchange of information between IEDs. Typically, these functions are not using human interaction and are performed via IEC 61850 publisher-subscriber communication. These are time-critical and the maximum accepted communication delay is in the range of several milliseconds. 

\subsubsection{Management of DERs} Increasing penetration of DERs at the consumer end adds a high degree of unpredictability to the power distribution grid's operation. Local distributed power generation from volatile DERs such as roof-top solar panels and small-scale wind turbines by prosumers (i.e. users who are able to both consume and produce electricity), challenges the flexibility of the power supply. The resulting bi-directional power flows complicate the protection mechanisms. Recently, microgrids have become an alternative to supplying power to isolated communities. A microgrid is a self-sufficient entity comprising of distributed generation, loads and local storage. For a flexible and reliable operation, proper control and automated protection system must prevail to overcome the challenges that emerge with DER integration in the islanded microgrids. 

\subsubsection{Demand response} The balance between the supply and demand of power is vital for the reliable operation of the power distribution grid, since storing energy in large scale is infeasible. This is largely dependent on the advanced metering infrastructure (AMI) based on smart meters which facilitates bi-directional communication between the power grid and the consumers. By careful observation of the smart meter parameters, suppliers and consumers can make wiser decisions such as peak demand saving.

\subsubsection{Managing big data} The addition of intelligent sensors, IEDs, PMUs and AMI enables the power distribution grid to acquire a massive amount of new information about different aspects of the grid. Data management and advanced analytics can support data-driven decision making and planning.

\subsection{Stakeholders}
The primary stakeholders involved in the power grid  fall under the different areas of power generation, transmission and distribution. However, in this paper, we focus only on the intents and SLAs between the key actors in the power distribution grid.

\begin{itemize}
    \item Distribution system operators (DSO): DSO is responsible for the operation of local electrical grid between power transmission system and the consumers. 
    \item Prosumers or customers: In smart grid context, the end users play a more active role by contributing towards the power generation and actively participating in management of electric power supply and demand.
    \item Demand response (DR) aggregator: DR aggregators create customized, automated controls for consumer loads and appliances that enable remote access while considering their preferences and behavioral patterns. DR aggregators facilitate the integration of demand-side technologies with ICT and AMI to encourage end-consumers and prosumers to participate in electricity markets.
    \item Communication service provider (CSP): The DSOs may demand different types of communication services from the CSP based on the characteristics of the different distribution grid operations.
\end{itemize}






The distribution grid operations impose stringent service requirements upon the ICT infrastructure. Table~\ref{table:1} categorizes them into the three major 5G use case categories: URLLC, mMTC and eMBB.


    

\begin{figure*}[ht!]
    \quad \quad \quad
    \includegraphics[height=6in, width=1\textwidth]{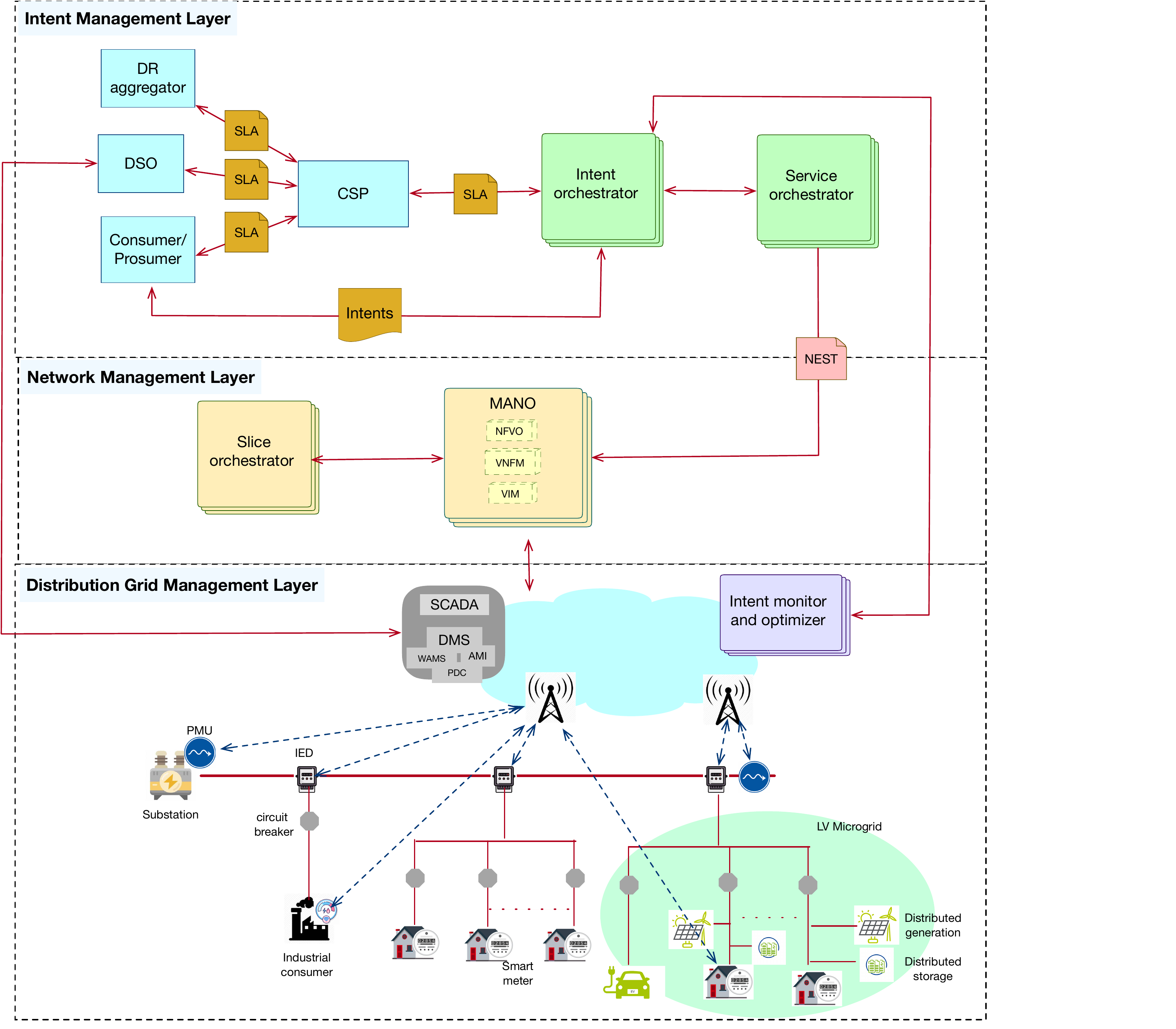}
    \caption{Proposed Intent-based network management and service  orchestration architecture.}
    \label{image: ibn_sg}
\end{figure*}

\section{Proposed Intent- and Network Slicing-Based
Framework}\label{sec:architect}
\begin{figure*}[ht!]
    \quad \quad \quad
    \includegraphics[height=6in, width=1\textwidth]{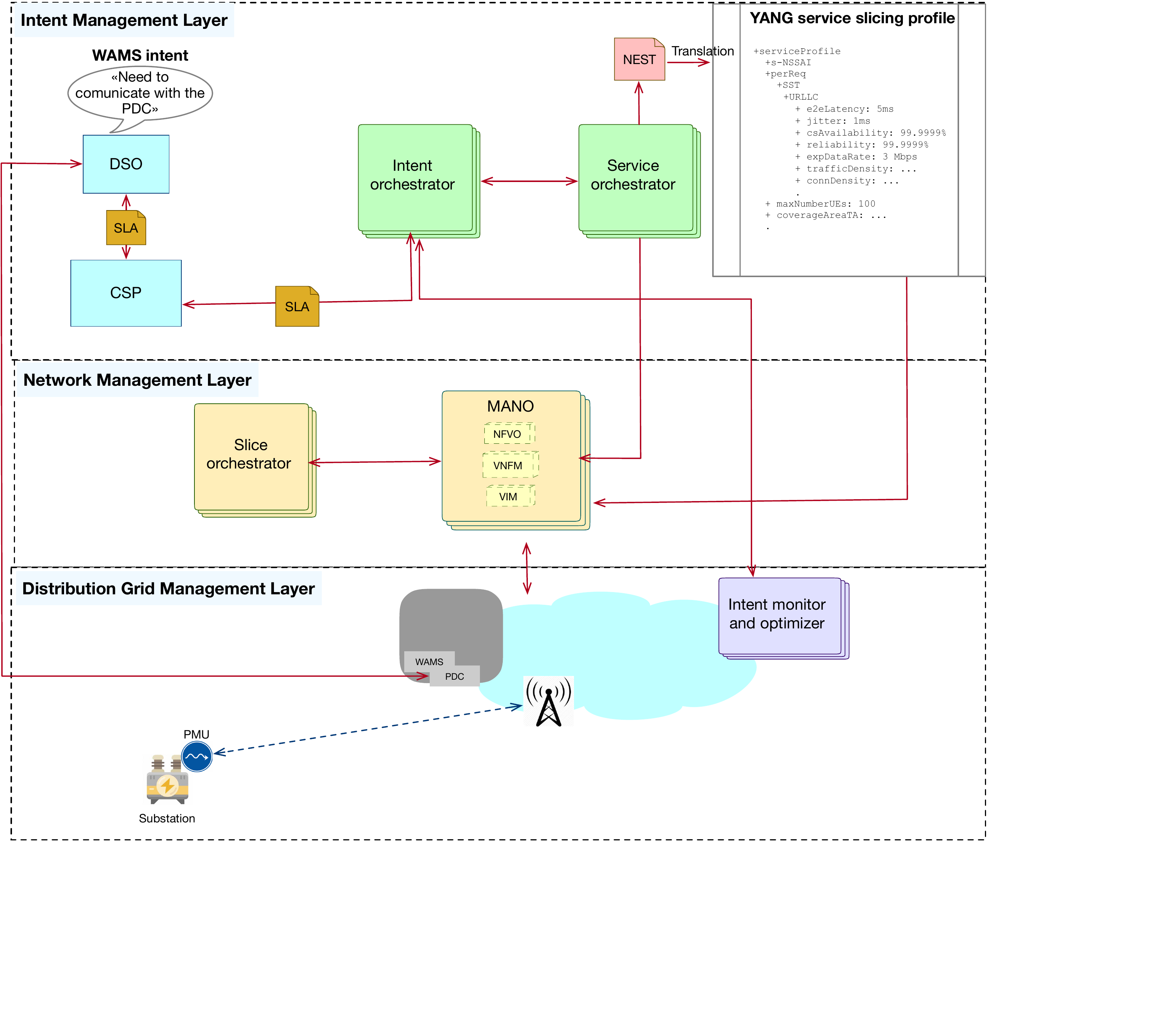}     \vspace{-3cm}
    \caption{An example of IBN WAMS application use case.}
    \label{image: ibn_example}
\end{figure*}

In this section, we propose an intent-based architecture for communication services in smart distribution grid given in Fig. \ref{image: ibn_sg} and explain it with one example use case in Fig. \ref{image: ibn_example}. The consumer/prosumer intents couple with the SLAs between the consumer/prosumer and CSP and the SLAs between the DSO and CSP, and the service provider follows them in order to provide the requested service creation and to orchestrate the framework. The proposed architecture is composed of three management layers consisting of intent, network, and distribution grid components. 

Note that this architecture can be interchangeably applied to any other vertical industry; thus, we also refer to it as a framework.

\subsection{Intent Management Layer}  \vspace{-0.1cm} 
The intent management layer is the critical part and provides the required functionality to implement the intent and  the service life-cycle. The intent processing starts with the information regarding SLAs from DSO and the grid consumer/prosumer relayed through the CSP. Grid users specify a high-level expectation for the network like \textit{connect residential consumers to the closest substation}. The intent orchestrator considers the SLA information and the users' intents during the intent translation and feasibility analysis of the required services. The intent translation consists of language processing by an intelligent module in order to iteratively produce the required service expectations for different grid users (see Table \ref{table:1}). A closed-loop optimization is performed with the intent monitor and optimizer module in the grid layer to ensure compliance of user intents with the configured service and network configuration. The intent monitor module also maintains an updated status of the resource utilization and availability in the grid management layer, which is utilized in the intent and service orchestration for the generation of feasible service profiles. The intent orchestrator cross validates the optimized service requirements with the assured SLAs for the users with the CSP and the DSO (Fig. \ref{image:intents}) before forwarding the service requirements to the service orchestrator. The service orchestration consists of generating a customer service model as per the provided service requirements from the intent orchestrator. Several feasibility checks are performed in order to ensure the required services are made available to the users, including the possible tradeoffs between different services and the dynamic nature of user requirements. A service model can be represented in the YANG modeling language \cite{service-model}, and following service orchestration, it instructs the network orchestrators to provision the required functionality for ensuring the generated service profiles for different user applications. The information exchange in the form of NEST profiles (Fig. \ref{image:intents}) happens between the service and slice orchestrators, followed by the network function provisioning through the management and network orchestrator.      

\subsection{Network Management Layer}
The middle layer consists of two functional blocks, namely, the MANO entity and the slice orchestrator. The slice orchestrator maintains the life-cycle (creation, update, deletion) of the network slices. The MANO entity manages and orchestrates the resources needed to realize network slice instances over the virtual network functions (VNFs)~\cite{etsi}. The NFV orchestrator (NFVO) manages NSIs; thus, it is responsible for the life-cycle management, fault, and performance management of NSIs. The VNF manager (VNFM) is responsible for the life-cycle (setting up, maintaining, and tearing down) of VNF instances. The virtualized infrastructure manager (VIM) connects to NFV infrastructures and sets up virtual links.   

\subsection{Distribution Grid Management Layer}
The bottom layer comprises the underlying physical infrastructure of the power distribution grid and 5G radio access network (RAN), and core network. The physical and logical components of the power distribution grid such as substations, PMUs, IEDs, circuit breakers, smart meters connect to the RAN in order to send the necessary information to the data aggregation points located in the proximity. These are then used by SCADA and DMS to execute the distribution grid automation processes.
The intent monitor and optimizer entity continuously observes the performance of the communication services delivered via the network slices. The observations are dispensed to the intent orchestrator, which makes a comparison between the desired and the observed KPIs and makes the adjustments accordingly. 
\subsection{Example}
We consider an example of the WAMS use case in Fig. \ref{image: ibn_example} requiring the establishment of network slices that deliver URLLC service. PMUs are responsible for the collection of phasor measurement data from the grid towards phasor data concentrators (PDC) and form the basis of the WAMS. An automated service provisioning is usually completed in the following steps:

\begin{enumerate}
    \item The interested service user (DSO) generates an intent \textit{"Need to communicate with central PDC"}, and enters in to an SLA.
    \item The intent orchestrator translates the intent to a service request with specific requirements and cross validates it with the SLAs. 
    \item The service request is forwarded to the service orchestrator that creates a service profile including the specifications for the GST and then a specific slice type is selected as NEST. The slice type selected for the WAMS use case is the URLLC and the slicing profile is generated by the service orchestrator for the slice orchestrator.
    \item The generated slicing profile is implemented via the network orchestrator (MANO) and the service provisioning life-cycle is completed for the requested service for the PMUs. 
\end{enumerate}

The intent monitor and optimizer, together with the intent orchestrator, dynamically adapt the KPIs to the network conditions and send updates to the service orchestrator. A new service slicing profile is generated likewise to accommodate the changing resource and network state in order to ensure the SLAs are satisfied for the users. The service slicing profiles consist of various parameters that are adjusted in order to enable the generated profiles to continuously meet the service requirements for the intent provided by the WAMS users.\par


\section{Concluding Remarks}\label{sec:con}
Automating the management and control of the power distribution grids provides benefits such as real-time operation of time-critical tasks, reduction of human intervention, thus, leading to reduced associated cost, high level of security of supply, increased productivity, resiliency, and dependability. In this paper, we proposed an architecture to facilitate the distribution grid automation using the dynamic and closed-loop mechanism of IBN and communication service provisioning using 5G network slicing. The architecture is service based consisting of distribution, network management and intent management layers. By providing one example, we illustrated the workflow from intent definition to slice realization in the distribution grid.

\section*{Acknowledgment}
This paper has been funded by CINELDI - Centre for intelligent electricity distribution, an 8 year Research Centre under the FME-scheme (Centre for Environment-friendly Energy Research, 257626/E20). The authors gratefully acknowledge the financial support from the Research Council of Norway and the CINELDI partners.

\bibliographystyle{ieeetr}
\typeout{}
\bibliography{main}

\end{document}